\documentclass[useAMS,usenatbib]{mn2e}
\usepackage{amsmath,amssymb,graphics,rotating,multirow,bm,color,hyperref}

\hypersetup{
    colorlinks=true,
    linkcolor=red,
    citecolor=green,
}

\title[Effects of Coupled Scalar Field on Structure Formation]
  {On the Effects of Coupled Scalar Fields on Structure Formation}
\author[Baojiu~Li, John~D.~Barrow]
  {Baojiu~Li$^{1,2,}$\thanks{E-mail: b.li@damtp.cam.ac.uk}, John~D.~Barrow$^{1,}$\thanks{E-mail: j.d.barrow@damtp.cam.ac.uk}\\
  $^1$DAMTP, Centre for Mathematical Sciences, University of Cambridge, Wilberforce Road, Cambridge CB3 0WA, UK\\
  $^2$Kavli Institute for Cosmology Cambridge, Madingley Road, Cambridge CB3 0HA, UK} 
\date{\today}

\pagerange{\pageref{firstpage}--\pageref{lastpage}} \pubyear{2010}

\def\LaTeX{L\kern-.36em\raise.3ex\hbox{a}\kern-.15em
    T\kern-.1667em\lower.7ex\hbox{E}\kern-.125emX}

\begin{document}

\label{firstpage}

\maketitle

\begin{abstract}
A coupling between a scalar field (representing the dark energy) and dark
matter could produce rich phenomena in cosmology. It affects cosmic
structure formation mainly through the fifth force, a velocity-dependent
force that acts parallel to particle's direction of motion and proportional
to its speed, an effective rescaling of the particle masses, and a modified
background expansion rate. In many cases these effects entangle and it is
difficult to see which is the dominant one. Here we perform $N$-body
simulations to study their qualitative behaviour and relative importance in
affecting the key structure formation observables, for a model with
exponential scalar field coupling. We find that the fifth force, a prominent
example of the scalar-coupling effects, is far less important than the
rescaling of particle mass or the modified expansion rate. In particular,
the rescaling of particle masses is shown to be the key factor leading to
less concentration of particles in halos than in $\Lambda$CDM, a pattern
which is also observed in previous independent coupled scalar field
simulations.
\end{abstract}


\section{Introduction}

\label{sect:intro}

The nature of the dark energy \citep{cst2006} driving an apparent
acceleration of the universe has been a cosmological puzzle for more than a
decade. Models incorporating scalar fields are the most popular proposal to
explain it, those, not only because of their mathematical simplicity and
phenomenological richness, but also because the scalar field is a natural
ingredient of many high-energy physics theories. A scalar field contributes
a single dynamical degree of freedom which can interact indirectly with
other matter species through gravity or couple directly to matter, producing
a fifth force on the matter which creates violations of the Weak Equivalence
Principle (WEP). This second possibility of direct coupling to matter was
introduced with the hope that such a coupling could potentially alleviate
the coincidence problem of dark energy \citep{amendola2000} and has since
then attracted a great deal of attention (see, for examples, \citet{bm2001,
amendola2004, koivisto2005, lln2006, bclm2008, bflt2008, bft2008,
bcclm2009} and references therein). It was also
investigated in the context of theoretical studies of the cosmological
variation of the fine structure constant \citep{sbm2002}.

If there is a direct coupling between the scalar field and baryons, then the
baryonic particles will experience a fifth force, which is severely
constrained by observations, unless there is some special mechanism to
suppress the fifth-force effects. This happens in chameleon models, where
the scalar field (the 'chameleon') gains mass in high-density regions (where
observations and experiments are performed) whereas the fifth force effects
are confined to undetectably small distances \citep{kw2004, ms2007}. A common approach which avoids such
complications is to assume that the scalar field couples only to the dark
matter, an idea seen frequently in models with a coupled dark sector (\emph{e.g.}~\citet{cms2009, vmm2010, sjp2010}). In this work our
scalar field will not be chameleon-like as this case has already been
investigated elsewhere \citep{lz2009, zmlhf2010, lz2010}.

A scalar field coupled to (dark) matter could affect cosmic structure
formation in various ways. Firstly, the background expansion rate gets
modified, which will lead to faster or slower clustering of matter
particles; secondly, the coupling effectively rescales the mass of the
particles for the coupled matter species, changing the source term of the
Poisson equation, which receives a further contribution from density
perturbations of the scalar field; thirdly, the coupling to the scalar field
produces a fifth force between matter particles, helping matter to cluster
more strongly; finally, there is an extra velocity-dependent force on the
coupled particles, which can be viewed either as a frictional force or as
part of the fifth force under a frame transformation, and this force, being
attractive, also promotes matter clustering.

Not all of these effects are always manifest. In fact, in some
coupled-scalar-field models one or more then one of them could be
negligible. An example is the model considered in \cite{lz2009, lz2010},
where only the third effect, \emph{i.e.}, the fifth force, is
non-negligible. This situation is also assumed in some other studies of the
effect of a fifth force on structure formation, \emph{e.g.}, the ReBEL model 
\cite{ngp2005}, where a Yukawa-type extra force is added while the
background cosmology is taken to be the same as $\Lambda$CDM. 

In other, more general, models, however, a coupling between scalar field
(dark energy) and (dark) matter often not only produces the fifth force, but
equally likely creates other effects listed above. Examples are the models
investigated in \cite{mqmab2004, bprs2010,lb2010}, where the modified
expansion rate, varying particle mass and frictional force are all
non-negligible. The situation then becomes complicated here, because these
could have both positive and negative effects on the structure formation,
which are difficult to disentangle.

In order to clarify the importance of all these effects, we have to make
detailed analysis by suppressing one or more of these effects and then
comparing the results -- which is our primary aim of this work, at least for
the model considered \cite{lb2010}. In particular, we would like to see
how those above effects affect the nonlinear matter power spectrum, mass
function and profiles of dark matter halos. The latter is quite interesting,
as it has been shown in ref. \cite{bprs2010, lb2010} that the coupling,
which is supposed to boost clustering of matter, does indeed suppress the
density in the inner region of the halos, possibly due to the frictional
force and/or the varying mass effects. A clarification of the importance of
these effects is also relevant to the general model tests. As we mentioned
above,  the majority of the investigations of the fifth force to date focus
only on the fifth force itself but neglect the other effects. Yet the same
physics responsible for the fifth force will often create other associated
effects and the latter must be taken into account for the sake of
consistency. If it turns out that the fifth force effects do not dominate
the others, then a model-independent test of the fifth force might be
difficult to obtain in cosmology, since different models predict very
different background expansion and mass variation.

Since we want to investigate the nonlinear regime, we will use the $N$-body
simulation technique introduced in \cite{lz2010, lb2010} and also applied in \cite{li2010, lmb2010a, lmb2010b}. Note that other
approaches to $N$-body simulations for (coupled) scalar field and related
models, without solving the scalar field equation of motion explicitly, have
been also used in various previous work, \emph{e.g.}, \cite{mqmab2004,
bprs2010, lj2003, mmbk2003, kk2006, sf2007,
fr2007, knp2009, knp2010, hj2009, hkk2010, baldi2010, bp2010, bv2010, hjv2010, ddempb2010}, while
the new feature of our approach is that we solve the scalar field equation
directly (in the quasi-static limit, for more details see \cite{lz2010,
lb2010}).

The outline of the paper is as follows: in Sect.~\ref{sect:eqn} we introduce
the basic equations needed to understand the underlying physics of the
model. In Sect.~\ref{sect:simu} we briefly describe the simulations we have
performed for the study of the different effects of a coupled scalar field
on structure formation, and then display and discuss the numerical results;
we summarise and conclude in Sect.~\ref{sect:con}.

\section{Basic Equations}

\label{sect:eqn}

All the equations relevant for the simulations used here are derived and
discussed in detail in \cite{lb2010}, but to make the present work
self-contained we list the minimum set necessary for us to understand the
physical evolution. Instead of writing down the field equations directly, we
start from a Lagrangian 
\begin{equation}
\mathcal{L}=\frac{1}{2}\left[ \frac{R}{\kappa }-\nabla ^{a}\varphi \nabla
_{a}\varphi \right] +V(\varphi )-C(\varphi )\mathcal{L}_{\mathrm{DM}}+
\mathcal{L}_{\mathrm{S}}\ \ 
\end{equation}
where $R$ is the Ricci scalar, $\kappa =8\pi G$ with $G$ the gravitational
constant, $\mathcal{L}_{\mathrm{DM}}$ and $\mathcal{L}_{\mathrm{S}}$ are
respectively the Lagrangian densities for dark matter and standard model
fields, $\varphi $ is the scalar field, and $V(\varphi )$ its potential; the
coupling function $C(\varphi )$ characterises the coupling between $\varphi $
and dark matter. Given $V(\varphi )$ and $C(\varphi )$ a model is then fully
specified.

Varying the total action with respect to the metric $g_{ab}$, we obtain the
following expression for the total energy-momentum tensor in this model: 
\begin{eqnarray}\label{eq:emt_tot}
T_{ab} &=& \nabla_{a}\varphi \nabla_{b}\varphi -g_{ab}\left[ \frac{1}{2}
\nabla^{c}\nabla _{c}\varphi -V(\varphi )\right]   \nonumber\\
&&+C(\varphi )T_{ab}^{\mathrm{DM}}+T_{ab}^{\mathrm{S}}
\end{eqnarray}
where $T_{ab}^{\mathrm{DM}}$ and $T_{ab}^{\mathrm{S}}$ are the
energy-momentum tensors for (uncoupled) dark matter and standard model
fields. The existence of the scalar field and its coupling change the form
of the energy-momentum tensor, and so modify the cosmology from background
expansion to structure formation.

The coupling to scalar field produces a direct interaction (a.k.a.~the fifth
force) between dark matter particles, due to the exchange of scalar quanta.
This is best illustrated by the geodesic equation for dark matter particles 
\begin{equation}\label{eq:geodesic}
\frac{d^{2}\mathbf{r}}{dt^{2}}=-\vec{\nabla}\Phi -\frac{C_{\varphi }(\varphi
)}{C(\varphi )}\vec{\nabla}\varphi 
\end{equation}
where $\mathbf{r}$ is the position vector, $t$ the (physical) time, $\Phi $
the Newtonian potential and $\vec{\nabla}$ is the spatial derivative. $C_{\varphi }=dC/d\varphi $. The second term on the right-hand side is the
fifth force and only exists for coupled matter species (dark matter in our
model). The fifth force also changes the clustering properties of the dark
matter. Note that on very large scales $\varphi $ is homogeneous and the
fifth force vanishes.

In order to solve the two equations above numerically we need to solve both
the time evolution and the spatial distribution of $\varphi $, and this
could be done using the scalar field equation of motion 
\begin{equation}
\nabla ^{a}\nabla _{a}\varphi +\frac{dV(\varphi )}{d\varphi }+\rho _{\mathrm{
DM}}\frac{dC(\varphi )}{d\varphi }=0
\end{equation}
or equivalently 
\begin{equation}
\nabla ^{a}\nabla _{a}\varphi +\frac{dV_{eff}(\varphi )}{d\varphi }=0
\end{equation}
where we have defined 
\begin{equation}
V_{eff}(\varphi )=V(\varphi )+\rho _{\mathrm{DM}}C(\varphi ).
\end{equation}
The background evolution of $\varphi $ can be solved easily once we know the
current $\rho _{\mathrm{DM}}$, because $\rho _{\mathrm{DM}}\propto a^{-3}$.
We can then divide $\varphi $ into two parts, $\varphi =\bar{\varphi}+\delta
\varphi $, where $\bar{\varphi}$ is the background value, and $\delta
\varphi $ the (not necessarily small and linear) perturbation, and subtract
the background scalar-field equation of motion from the full equation to
obtain the equation of motion for $\delta \varphi $. In the quasi-static
limit where we can neglect time derivatives of $\delta \varphi $ compared
with its spatial derivatives (which turns out to be a good approximation for
our simulations, because the simulation box is much smaller than the
observable Universe), we get 
\begin{equation}\label{eq:scalar_eom_pert}
\vec{\nabla}^{2}\varphi =\frac{dC(\varphi )}{d\varphi }\rho _{\mathrm{DM}}-%
\frac{dC(\bar{\varphi})}{d\bar{\varphi}}\bar{\rho}_{\mathrm{DM}}+\frac{%
dV(\varphi )}{d\varphi }-\frac{dV(\bar{\varphi})}{d\bar{\varphi}}\ \ 
\end{equation}%
where $\bar{\rho}_{\mathrm{DM}}$ is the background dark-matter density.

Once $\rho_{\mathrm{DM}}$ is known on a grid, we can then solve $\delta
\varphi $ on that grid using a nonlinear Gauss-Seidel relaxation method (in
our simulations we have modified \texttt{MLAPM} \cite{kgb2001}, a
publicly available $N$-body code using a self-adaptive refined grid so that
high resolutions can be achieved in high-density regions). Since $\bar{%
\varphi}$ is also known, we can then obtain the full solution of $\varphi =%
\bar{\varphi}+\delta \varphi $. This completes the computation of the source
term for the Poisson equation: 
\begin{equation}\label{eq:Poisson}
\vec{\nabla}^{2}\Phi =\frac{\kappa }{2}\left[ C(\varphi )\rho _{\mathrm{DM}%
}-C(\bar{\varphi})\bar{\rho}_{\mathrm{DM}}+\delta \rho _{\mathrm{B}}-2\delta
V(\varphi )\right] ,
\end{equation}
where $\delta \rho _{\mathrm{B}}\equiv \rho _{\mathrm{B}}-\bar{\rho}_{%
\mathrm{B}}$ and $\delta V(\varphi )\equiv V(\varphi )-V(\bar{\varphi})$ are
respectively the density perturbations of baryons and scalar field (note
that we have neglected perturbations in the kinetic energy of the scalar
field because they are always very small for our model).

We can then solve Eq.~(\ref{eq:Poisson}) using a linear Gauss-Seidel
relaxation method on the same grid to obtain $\Phi $. With both $\Phi $ and $\varphi$ in hand, Eq.~(\ref{eq:geodesic}) can then be used to compute the
forces on the dark matter particles, and once we have the forces, we can
perform all the standard $N$-body operations such as momentum-kick,
position-drift, time-stepping and so on.

Eqs.~(\ref{eq:emt_tot} - \ref{eq:Poisson}) are all what we need to complete
an $N$-body simulation for coupled scalar field cosmology \cite{lb2010},
and from them we can see where the effects of the scalar-coupling enter:

\begin{enumerate}
\item The modified background expansion rate mainly affect
the particle movements and time-stepping, \emph{i.e.}, Eq.~(\ref{eq:geodesic}), because in the simulations we use the cosmic scale
factor $a$, instead of $t$, as the time variable and $d/dt=\dot{a}d/da$. 

\item The varying mass effect is seen directly from Eq.~(\ref{eq:Poisson}), which shows that the contribution of $\rho_{\mathrm{DM}}$ to the source
term of the Poisson equation is normalised by $C(\varphi )$ which is
different from 1 in general. In our model it is not true that the mass
of dark matter particles is really varying, but the net effect is just
equivalent to such a variation. 

\item The fifth force appears explicitly on the right-hand side of Eq.~(\ref{eq:geodesic}), but is only for coupled matter species (dark matter). 

\item The velocity-dependent (or frictional) force 
hides in the fact that Eqs.~(\ref{eq:geodesic}, \ref{eq:scalar_eom_pert})
are given in \emph{different} gauges: Eq.~(\ref{eq:geodesic}) is
the force for a dark-matter particle and is given in that particle's rest
frame, while Eq.~(\ref{eq:scalar_eom_pert}) is written in the fundamental
observer's frame. As a result, in order to use the $\delta \varphi $ solved
from Eq.~(\ref{eq:scalar_eom_pert}) in Eq.~(\ref{eq:geodesic}), we need to
perform a frame transform $\vec{\nabla}\delta \varphi \rightarrow \vec{\nabla%
}\delta \varphi +a\dot{\bar{\varphi}}\dot{\mathbf{x}},$ where $\dot{\mathbf{x%
}}$ is the comoving velocity of the particle relative to the fundamental
observer. This force is therefore expressed as $-\frac{C_{\varphi }}{C}a\dot{%
\varphi}\dot{\mathbf{x}}$, and obviously the faster a particle travels the
stronger such force it feels.
\end{enumerate}

\section{Simulations and Results}

\label{sect:simu}

\subsection{Model and Simulation Details}

As is mentioned above, a coupled scalar field model is fully specified given
the exact forms of the bare potential $V(\varphi )$ and coupling function $C(\varphi )$. Here we shall choose the same model as \cite{lb2010}, with
an inverse power-law potential 
\begin{equation}
V(\varphi )=\frac{\Lambda ^{4}}{\left( \sqrt{\kappa }\varphi \right)
^{\alpha }}
\end{equation}
and the exponential coupling 
\begin{equation}
C(\varphi )=\exp \left( \gamma \sqrt{\kappa }\varphi \right) ,
\end{equation}
where $\Lambda $ is a constant with mass dimension and $\Lambda ^{4}$ is of
order the dark-energy density today; $\alpha ,\gamma $ are dimensionless
parameters. We choose $\alpha =0.1$ so that the potential is flat enough to
enable a slow-roll of $\varphi $ (which accounts for the dark energy); for $\gamma $, we choose $|\gamma |\sim \mathcal{O}(0.1)$ and $\gamma <0$ so that
both $V(\varphi )$ and $V_{eff}(\varphi )$ are of runaway type\footnote{Since $V_{eff}(\varphi )$ is of runway type there is nothing to stop the
scalar field rolling down $V_{eff}$, so typically we shall have $\sqrt{\kappa }\varphi \sim \mathcal{O}(1)$ today, which makes $C(\varphi)$
deviate significantly from 1. Increasing $|\gamma |$ will makes this problem
more severe, and this is why we set $|\gamma |\sim \mathcal{O}(0.1)$ rather
than $\mathcal{O}(1)$.}.

\cite{lb2010} have given a very detailed description of the
technicalities of the $N$-body simulations for coupled scalar-field
theories, and so we shall not repeat this methodology here. Roughly
speaking, the most important distinction between our simulation and others
is that we have solved the scalar field equation of motion explicitly on a
grid (in the quasi-static limit). Because of this, we have been able to
solve the fifth force (and the frictional force) numerically without
recourse to analytical approximations. Furthermore, we have incorporated
both \emph{time} and the \emph{space} variations of the particle mass (or
more rigorously of $C(\varphi )$) because we have spatial information about
the scalar field distribution.

Since our aim is to test the significance of each of the four above-named
effects, we choose to suppress one of them at one time. Together with the
full model, where all effects are included, and the $\Lambda $CDM model for
comparison, we then have six models to simulate. Furthermore, we consider
two different choices of the coupling strength $\gamma $. So in total we
have 11 models, details of which are summarised in the following table: 
\begin{table*}
\begin{minipage}{170mm}
\begin{tabular}{cccccccccc}
\hline\hline
simulation no. & $\alpha$ &&&& $\gamma$ &&&& simulation description \\
\hline
L   & 0.0 &&&& 0.00 &&&& pure $\Lambda$CDM \\
S1  & 0.1 &&&& $-0.10$ &&&& full coupled scalar field \\
S1a & 0.1 &&&& $-0.10$ &&&& scalar field with the frictional force $-\frac{C_\varphi}{C}a\dot{\bar{\varphi}}\dot{\mathbf{x}}$ suppressed \\
S1b & 0.1 &&&& $-0.10$ &&&& scalar field with the fifth force $-\frac{C_\varphi}{C}\vec{\nabla}\delta\varphi$ suppressed \\
S1c & 0.1 &&&& $-0.10$ &&&& scalar field with the (time \emph{and} spatial) variation of mass $C(\varphi)$ removed \\
S1d & 0.1 &&&& $-0.10$ &&&& scalar field with a $\Lambda$CDM background expansion \\
S2  & 0.1 &&&& $-0.20$ &&&& full coupled scalar field \\
S2a & 0.1 &&&& $-0.20$ &&&& scalar field with the frictional force $-\frac{C_\varphi}{C}a\dot{\bar{\varphi}}\dot{\mathbf{x}}$ suppressed \\
S2b & 0.1 &&&& $-0.20$ &&&& scalar field with the fifth force $-\frac{C_\varphi}{C}\vec{\nabla}\delta\varphi$ suppressed \\
S2c & 0.1 &&&& $-0.20$ &&&& scalar field with the (time \emph{and} spatial) variation of mass $C(\varphi)$ removed \\
S2d & 0.1 &&&& $-0.20$ &&&& scalar field with a $\Lambda$CDM background expansion \\
\hline \hline
\end{tabular}
\end{minipage}
\end{table*}

The physical parameters we adopt in all simulations are as
follows: the present-day dark-energy fractional energy density $\Omega_{\mathrm{DE}}=0.743$ and $\Omega _{m}=\Omega _{\mathrm{CDM}}+\Omega _{\mathrm{%
B}}=0.257$, $H_{0}=71.9$~km/s/Mpc, $n_{s}=0.963$, $\sigma _{8}=0.761$. The size of
simulation box is $64h^{-1}$~Mpc with $h=H_{0}/(100~\mathrm{km/s/Mpc})$. In all these simulations, the mass resolution is $1.114\times
10^{9}h^{-1}~M_{\bigodot }$, the particle (both dark matter and baryons)
number is $256^{3}$, the domain grid
is a $128\times 128\times 128$ cubic and the finest refined grids have 16384
cells on each side, corresponding to a force resolution of order $12h^{-1}~$kpc.

\begin{figure*}
\centering \includegraphics[scale=1.05] {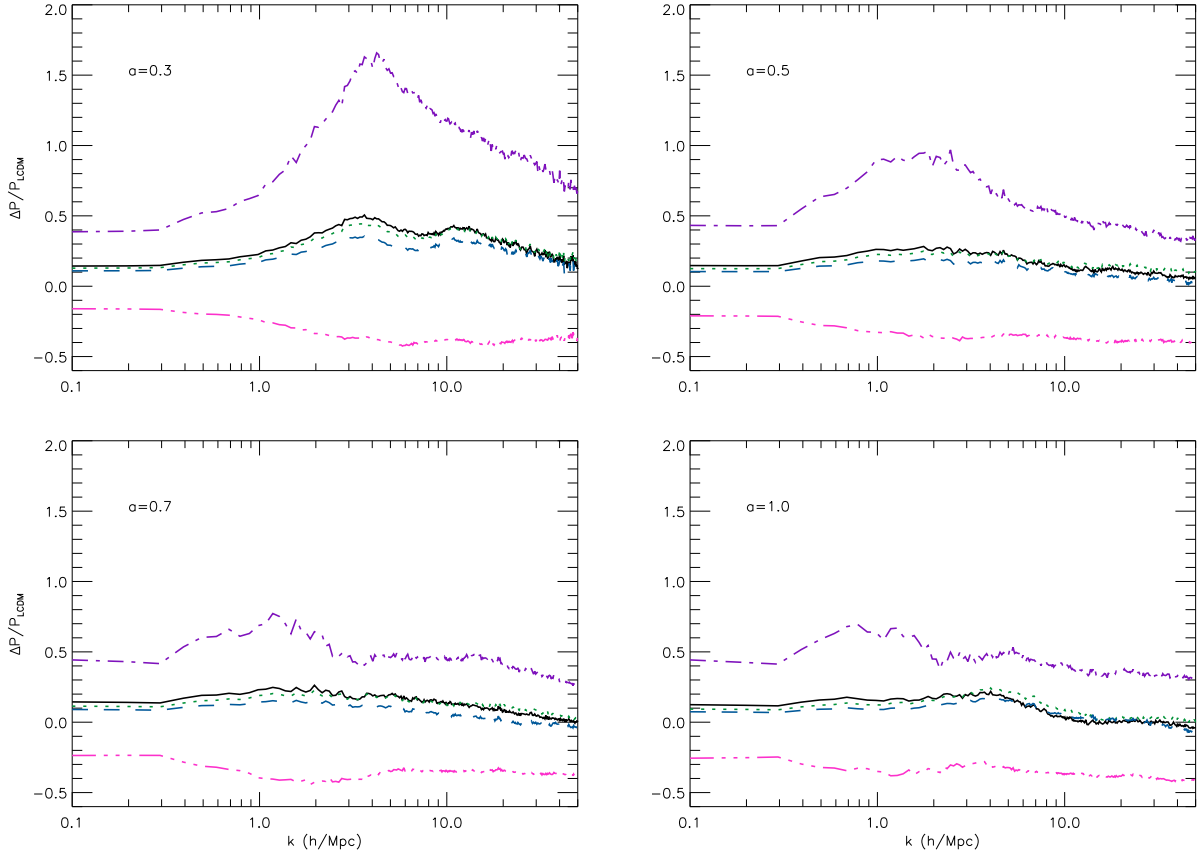}
\caption{(Colour Online) Fractional changes of the nonlinear matter power
spectrum with respect to the $\Lambda $CDM result at four different output
times: $a=0.3$ (\emph{upper left panel}), $a=0.5$ (\emph{upper right panel}), $a=0.7$ (\emph{lower left panel}) and $a=1.0$ (\emph{lower right panel}).
In each panel the results of the simulations S1 (full coupled scalar
simulation), S1a (frictional force suppressed), S1b (fifth force
suppressed), S1c (mass variation removed) and S1d ($\Lambda $CDM background)
are represented  by the black solid, green dotted, blue dashed, purple
dash-dot and pink dash-dot-dot-dot curves respectively.}
\label{fig:power_s1}
\end{figure*}

\begin{figure*}
\centering \includegraphics[scale=1.05] {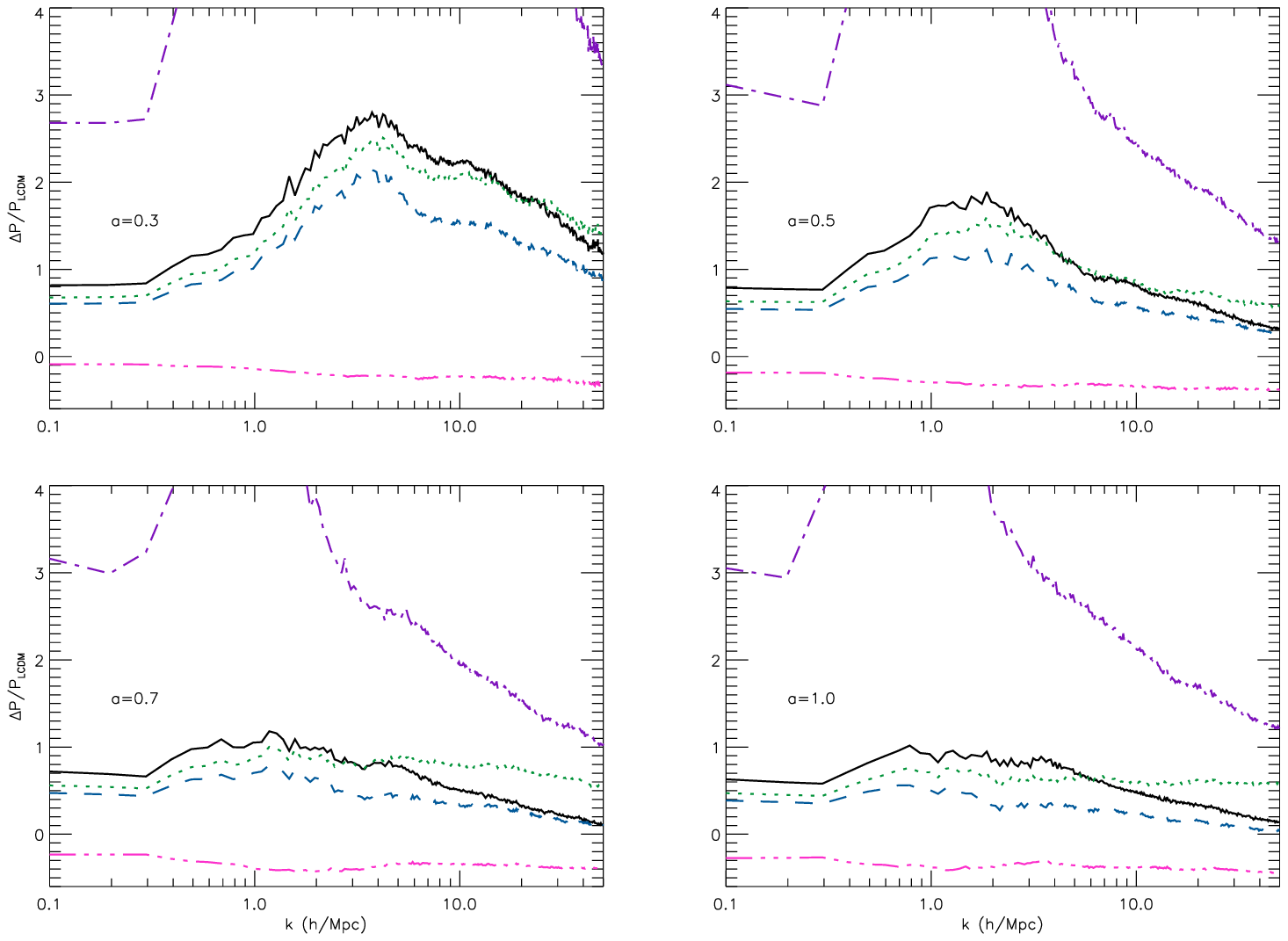}
\caption{(Colour Online) The same as Fig.~\protect\ref{fig:power_s1}, but for
the models where $\protect\gamma =-0.20$, \emph{i.e.}, S2 and S2a-d.}
\label{fig:power_s2}
\end{figure*}

\subsection{Numerical Results}

For the numerical results, we display the matter power spectrum, mass
function and halo density profiles for the 11 runs described above, and
discuss how they are affected by the individual effects from the scalar
coupling.

Before going to the details, it is helpful to have a quick browse about the
scalar-coupling effects: (I) The velocity-dependent force $-\frac{C_\varphi}{%
C}a\dot{\bar{\varphi}}\dot{\mathbf{x}}$ is parallel to the direction of
motion ($\dot{\mathbf{x}}$); because $\dot{\bar{\varphi}}>0$ and $\gamma<0$,
so it accelerates the particles. (II) The fifth force $-\frac{C_\varphi}{C}%
\vec{\nabla}\delta\varphi$ in this model is found \citep{lb2010} to be
parallel to gravity and the ratio between the magnitudes of the two is $%
2\gamma^2$ to a high precision; as such the fifth force both accelerates the
particles and increases their mutual attraction. (III) $C(\varphi)<1$
because $\gamma<0$ and $\varphi>0$, so that the contribution of dark matter
density to the source of Poisson equation gets weakened, effectively reducing
the gravity force and causing less mutual attraction between particles and
less clustering. (IV) For the chosen model and physical parameters, the
background expansion rate decreases as $|\gamma|$ increases \citep{lb2010},
making matter particles less diluted and cluster more. These facts are
important to bear in mind for discussions below.

\subsubsection{Matter Power Spectrum}

In Fig.~\ref{fig:power_s1} we have plotted the fractional change of the
nonlinear matter power spectrum with respect to $\Lambda $CDM prediction for
the simulations S1 and S1a-d. Roughly,  the deviation from the black solid
curve (S1) indicates the importance of a given coupled-scalar effect:
the larger the deviation is, the more that specific effect contributes to
the full coupled scalar field result. Of least importance is the
velocity-dependent force $-\frac{C_{\varphi }}{C}a\dot{\bar{\varphi}}\dot{\mathbf{x}}$ (green dotted curves). As this force accelerates particles, it
makes particles collapse faster to the regions of high density, and thus
slightly enhances the clustering. Consequently, suppressing it will decrease $P(k)$. Of the second least importance is the fifth force term $-\frac{C_{\varphi }}{C}\vec{\nabla}\delta \varphi $ (blue dashed curves). It not
only accelerates particles (towards the high-density regions) but also
increases the central force. As such its effect is also to enhance the
clustering of matter and neglecting it leads to smaller $P(k)$ on all scales.

Next is the varying mass effect (purple dash-dot curves), the presence
of which reduces the source of the Poisson equation and thus weakens
gravity. Obviously its effect is to produce weaker clustering of matter
particles and dropping it will increase $P(k)$ significantly.

he most important coupled-scalar effect comes from the modified background
expansion rate (pink dash-dot-dot-dot curves). As mentioned above, if the
universe expands more slowly (as in the case of our coupled scalar field
model \cite{lb2010}), then particles are less diluted and have more time to
cluster, resulting in a larger $P(k)$. Changing the background expansion to $\Lambda $CDM (\emph{i.e.}, increasing it) simply produces a smaller $P(k)$
than the full simulation. Note that the pink curves are consistently below
zero, indicating that although simulation S1d uses the same background
expansion rate as simulation L, $P(k)$ is smaller than for the latter, a
result that is again due to the fact that in S1d the varying mass effect is
taken into account, weakening the matter clustering and decreasing $P(k)$
(as discussed above, the fifth force does have the opposite effect but
cannot overcome this).

In Fig.~\ref{fig:power_s2} we have shown the same plots, but for the models
S2 and S2a-d. All the above analysis still applies but the effects just
become stronger. Interestingly, the most important coupled-scalar effects
(at least for our models, which are typical ones) are not those of the fifth
force, but are instead the modified background expansion rate and varying
mass of matter particles. This is understandable, because the magnitude of
the fifth force is about $2\gamma^{2}$ times that of gravity, and for our
S1 and S2 simulations the values are $0.02$ and $0.08$ respectively, while at the
same time the deviations of $C(\varphi)$ from 1 for these two models are $\sim 0.1$ and $\sim 0.3$ \cite{lb2010} and so are far stronger.

This means that one must be cautious about adding a Yukawa-type fifth force
to the $N$-body simulation, while keeping all other things the same as in $\Lambda $CDM, because the fifth force often introduces associated effects
which are more important

\begin{figure}
\centering \includegraphics[scale=0.5] {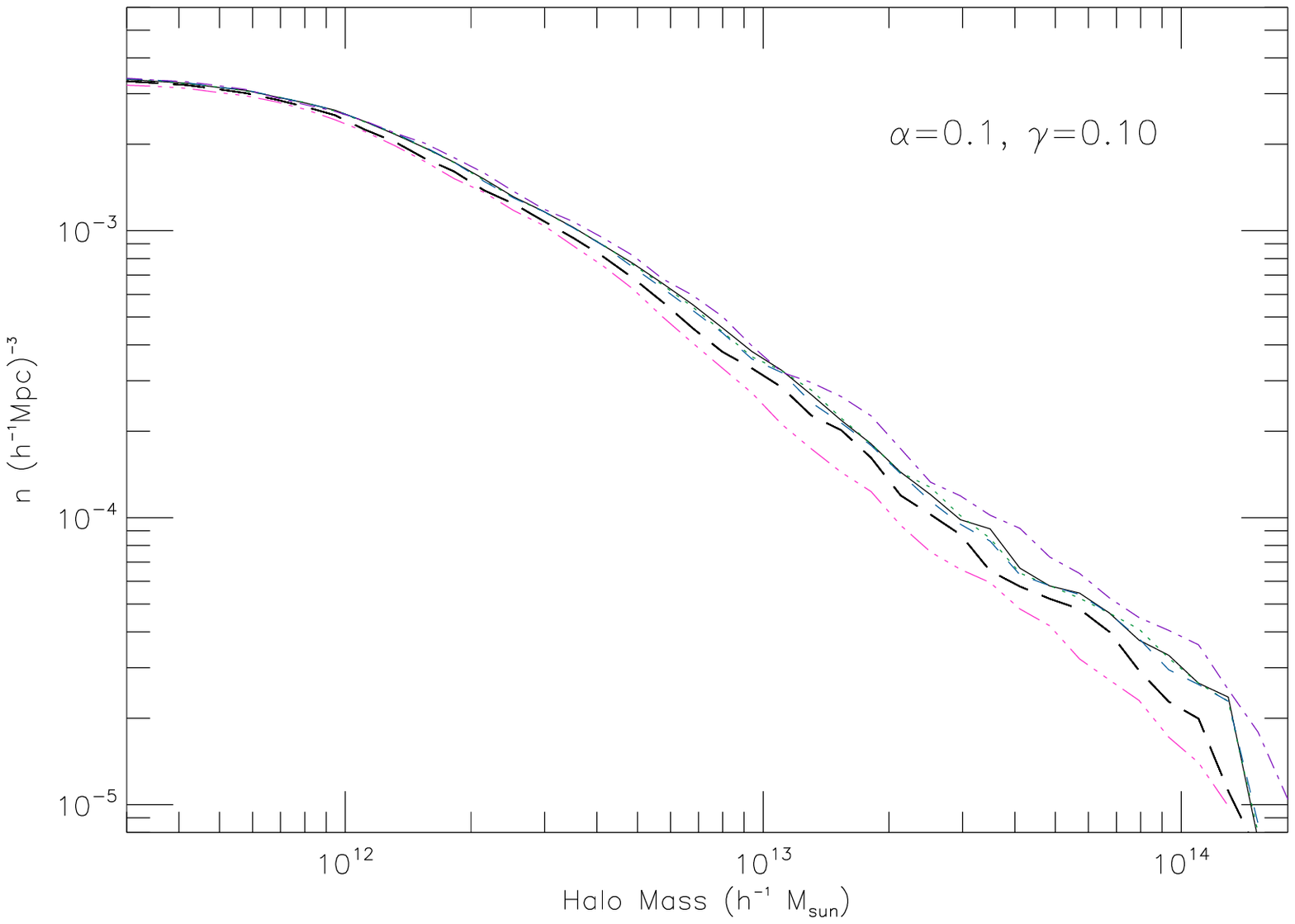}
\caption{(Colour Online) The mass functions for the models S1 (black solid
curve), S1a (green dotted), S1b (blue dashed), S1c (purple dash-dot) and S1d
(pink dash-dot-dot-dot) compared to the $\Lambda $CDM result (thick long
dashed curve). The horizontal axis is the virial mass of halos, in unit of $h^{-1}M_{\bigodot }$; the vertical axis is the halo number density in the
simulation box, in units of $\left( h^{-1}\mathrm{Mpc}\right) ^{-3}$.}
\label{fig:mf_s1}
\end{figure}

\begin{figure}
\centering \includegraphics[scale=0.5] {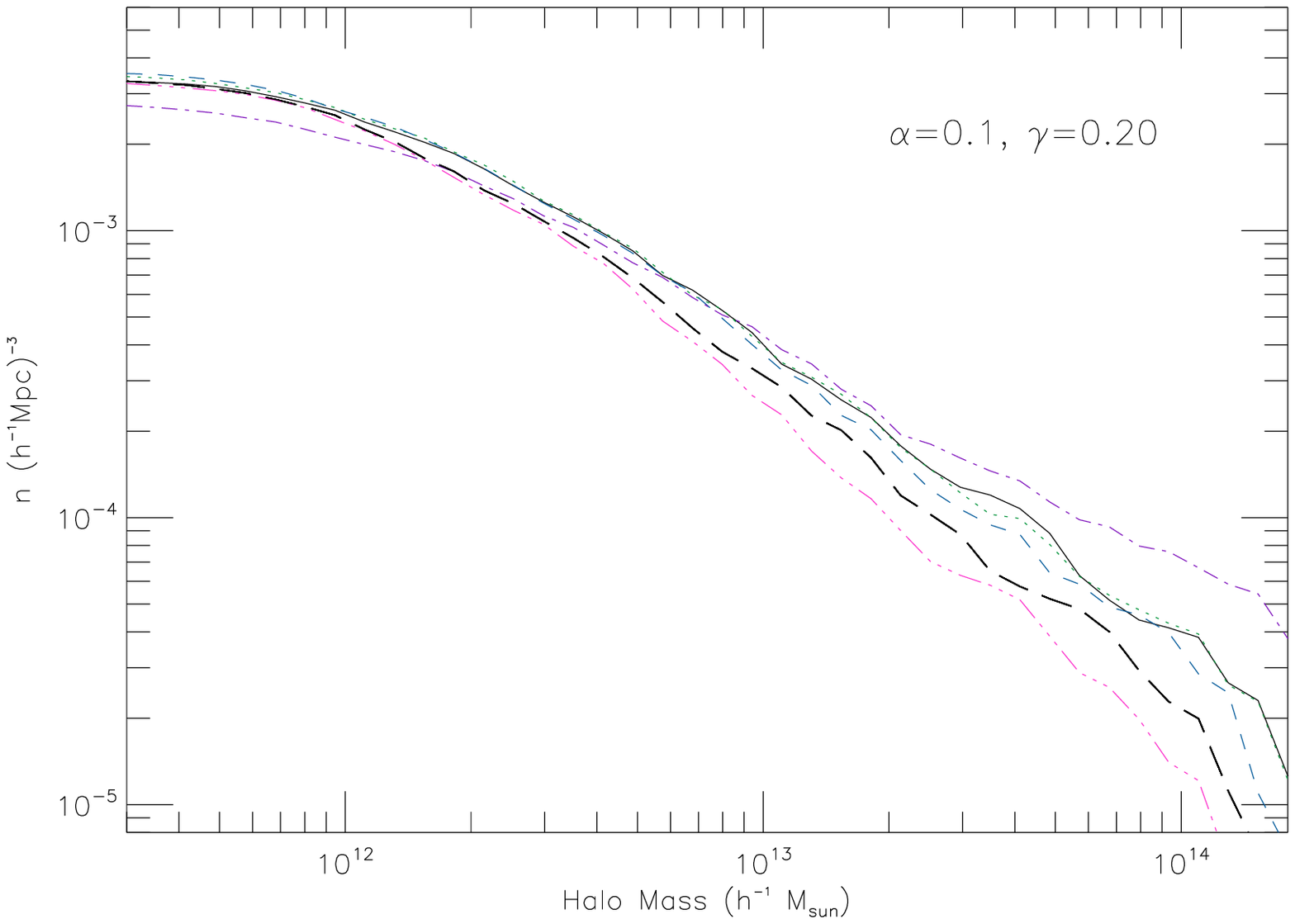}
\caption{(Colour Online) The same as Fig.~\protect\ref{fig:mf_s1}, but for
models S2 and S2a-d.}
\label{fig:mf_s2}
\end{figure}

\begin{figure}
\centering \includegraphics[scale=0.5] {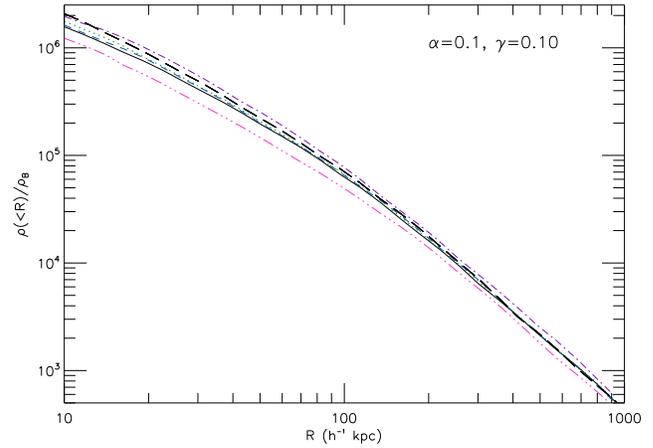}
\caption{(Colour Online) The internal density profile of the most massive
halo in the simulation box for the models S1 (black solid curve), S1a (green
dotted), S1b (blue dashed), S1c (purple dash-dot) and S1d (pink
dash-dot-dot-dot) compared to the $\Lambda $CDM result (thick long dashed
curve). The horizontal axis is the radius from halo centre, in unit of $h^{-1}\mathrm{Kpc}$, and the vertical axis is the overdensity.}
\label{fig:pf_s1}
\end{figure}

\begin{figure}
\centering \includegraphics[scale=0.5] {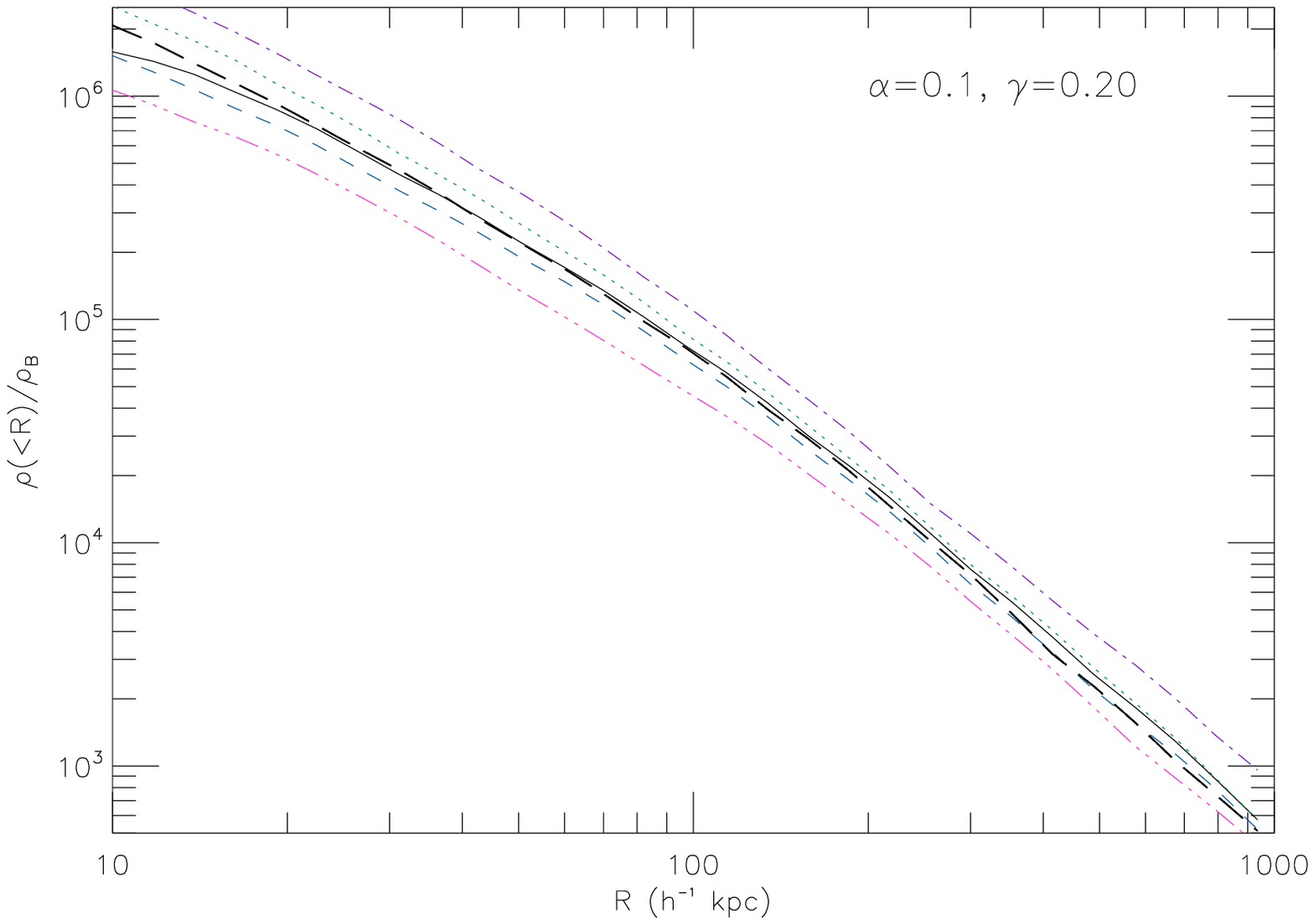}
\caption{(Colour Online) The same as Fig.~\protect\ref{fig:pf_s1}, but for
models S2 and S2a-d.}
\label{fig:pf_s2}
\end{figure}

\subsubsection{Mass Functions}

Fig.~\ref{fig:mf_s1} shows the mass functions of the simulations S1 and S1
compared with that of $\Lambda $CDM (L). Again, the deviation from the full
simulation result (black solid curve) indicates the order of importance of
the individual effects. The two least important factors are once more the
velocity-dependent force $-\frac{C_{\varphi }}{C}a\dot{\bar{\varphi}}\dot{\mathbf{x}}$ and the fifth force $-\frac{C_{\varphi }}{C}\vec{\nabla}\delta
\varphi $, both of which, according to our above analysis, enhance matter
cluster: suppressing them causes less clustering of matter and smaller mass
functions. Their influences however are quite weak, in particular that of
the velocity-dependent force.

The second important effect is the variation of mass or $C(\varphi )$. In case $C(\varphi )<1$ then, as mentioned above, the mutual gravitational
interaction between particles becomes weaker and particles will cluster less
strongly. As a result, removing this effect enhances the matter clustering
and leads to massive halos being created in larger abundance.

\begin{figure}[tbp]
\centering \includegraphics[scale=0.5] {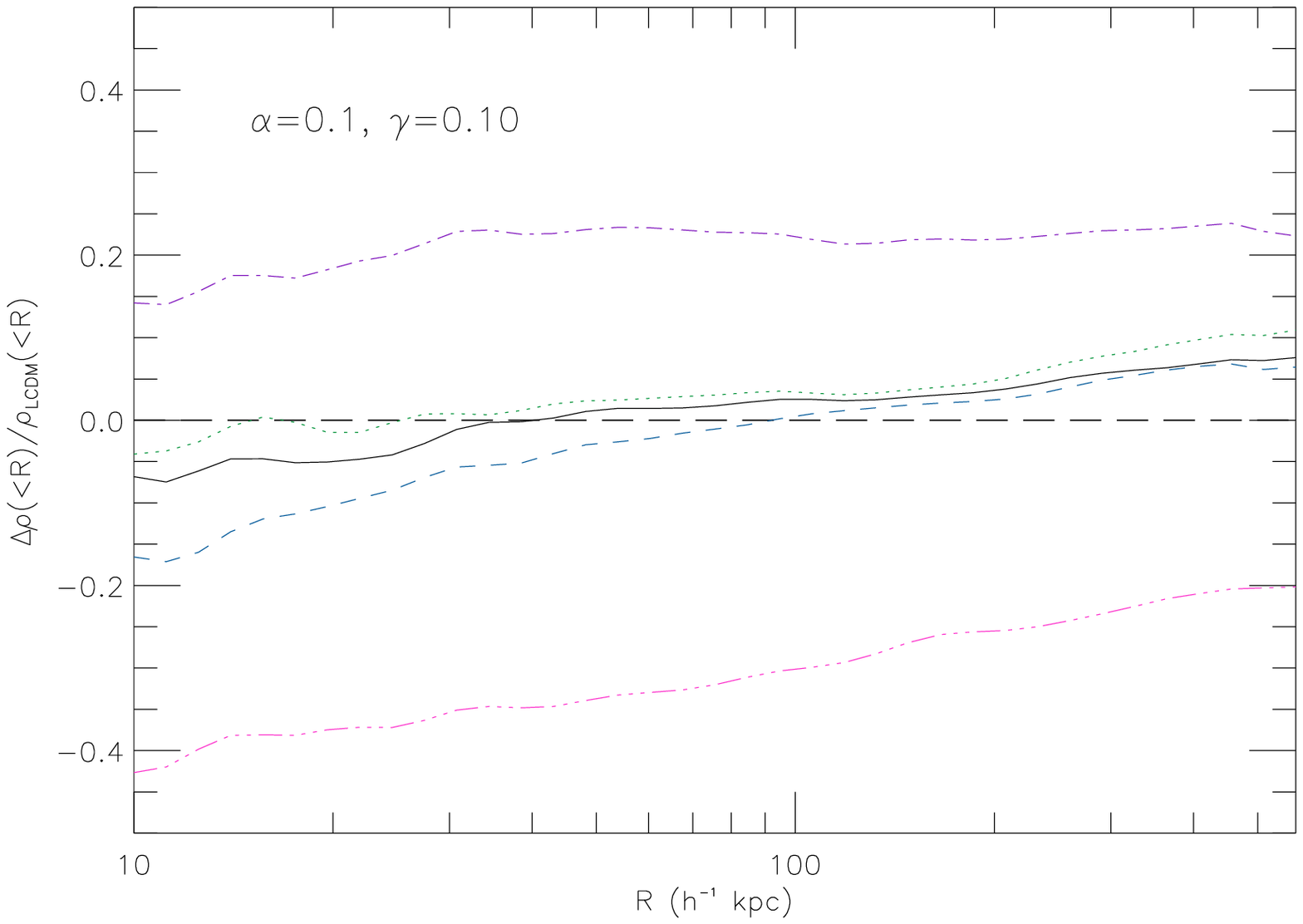}
\caption{(Colour Online) The fractional change of averaged halo density
profile for the models S1 (black solid curve), S1a (green dotted), S1b (blue
dashed), S1c (purple dash-dot) and S1d (pink dash-dot-dot-dot) compared to
the $\Lambda $CDM result (thick long dashed curve). The horizontal axis is
the distance from halo centre, in unit of $h^{-1}\mathrm{Kpc}$, and the
vertical axis is the fractional change of density with respect to the $\Lambda $CDM prediction.}
\label{fig:pfr_s1}
\end{figure}

The most influential effect from a coupled scalar field is the
modified background expansion. Changing it to a $\Lambda$CDM background
(which is faster) significantly underestimates the mass function, because
particles get more diluted and have less time to clump.

The corresponding results for models S2 and S2a-d are summarized in Fig.~\ref{fig:mf_s2}, and they show the same qualitative trend as Fig.~\ref{fig:mf_s1}
but the effects are just stronger, due to the stronger coupling $|\gamma |$
and thus more dramatic evolution of the scalar field $\varphi$ (which means that $C(\varphi)$ deviate more from unity and the expansion rate is decreased more compared to simulation L).

Although the mass function does show the imprint from a fifth force, this is
by no means unique and could easily be dominated over by the associated
coupled scalar effects like modified background expansion rate or varying
particle mass.

\begin{figure}
\centering \includegraphics[scale=0.5] {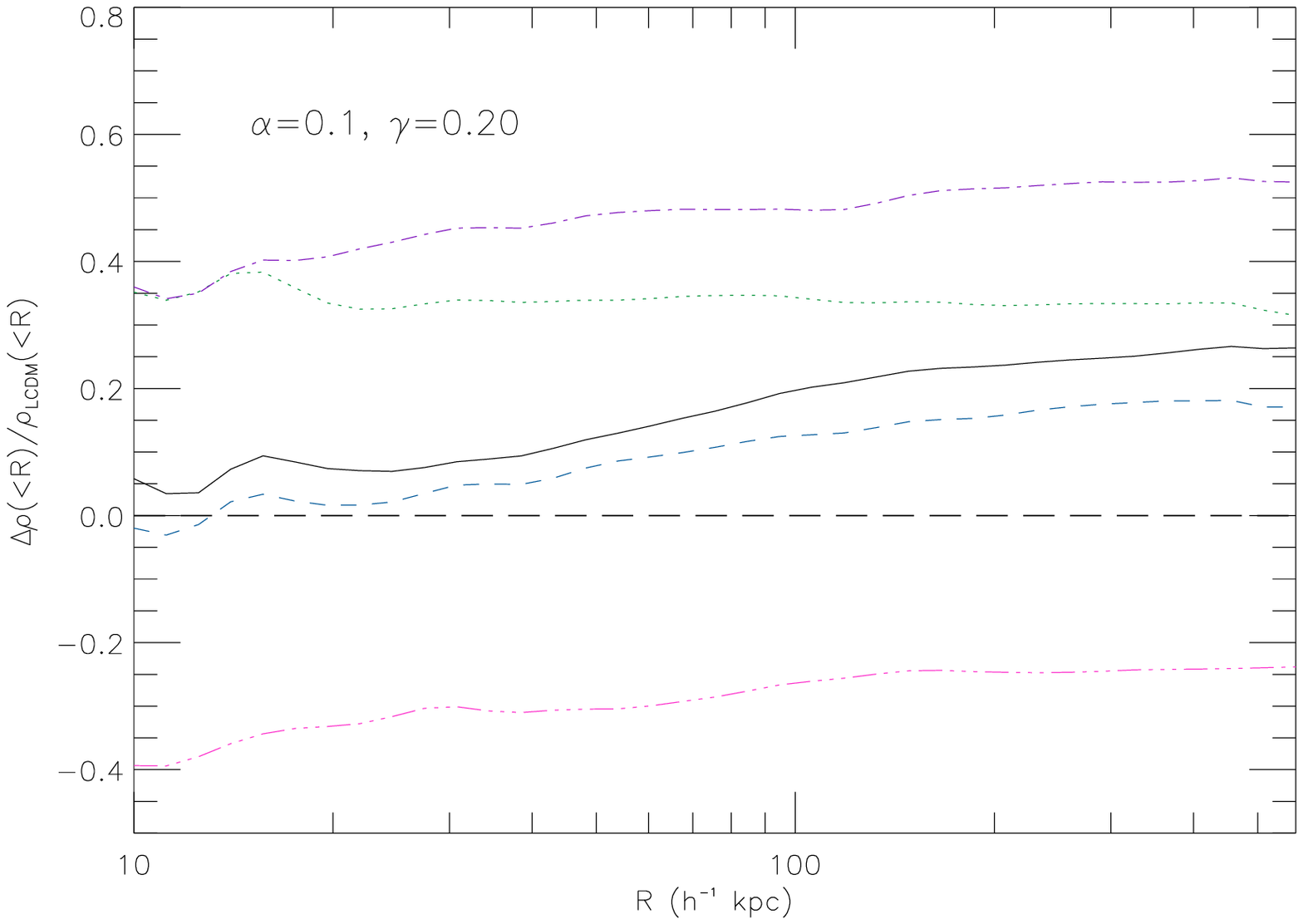}
\caption{(Colour Online) The same as Fig.~\protect\ref{fig:pfr_s1}, but for
models S2 and S2a-d.}
\label{fig:pfr_s2}
\end{figure}

\subsubsection{Halo Density Profiles}

Internal density profiles of the dark matter halos are another area where
the scalar field coupling could leave interesting imprints. For example, 
\cite{lz2010} showed that for a chameleon-like scalar field model the
density profiles could be either increased or decreased significantly by the
scalar coupling, depending on the environment of the halos. For
the models considered here, \citet{bprs2010, lb2010} have given
convincing evidence that the internal density profile has similar shape to $\Lambda $CDM, but could be somewhat suppressed in the very inner parts.

As an explicit example, Fig.~\ref{fig:pf_s1} shows the density profiles of
the most massive halo from each box for simulations L, S1 and
S1a-d. The suppression of the full simulation result (black solid curve)
compared with the $\Lambda $CDM prediction (thick solid dashed curve) is
evident below $R\sim 400h^{-1}\mathrm{Kpc}$.

If the velocity-dependent force $-\frac{C_{\varphi }}{C}a\dot{\bar{\varphi}}\dot{\mathbf{x}}$ is given up (green dotted curve), then the suppression is
moderated. As discussed in \cite{bprs2010}, this is because this force
effectively accelerates particles, making them travel faster and so
increases the total kinetic energy of the particles inside the halo. As a
result, removing this force will leave the particles with less kinetic
energy, meaning that they more easily fall towards the halo centre.
Unlike in \cite{bprs2010}, here this effect is not the major one (see
below).

The fifth force $-\frac{C_{\varphi }}{C}\vec{\nabla}\delta \varphi $ is
slightly more complicated. On one hand, it speeds the particles up,
increasing their kinetic energy; on the other, it enhances the mutual
attraction between particles, increasing the (magnitude of the) potential
energy of the halo. These two effects are just opposite, as deepening the
potential of a halo will pull more particles towards the centre and thus
increase the concentration. For simulation S1, it seems that the former
effect takes over, and dropping the fifth force (blue dashed curve) simply
decreases the kinetic energy of particles and make them more concentrated
towards the halo centre.

The effect of the varying-mass factor $C(\varphi )<1$ is unambiguously to
suppress the over-density inside the halos, as it decreases the source of the
Poisson equation and so weakens gravity, making the potential shallower.
Dropping it simply deepens the potential and attracts more particles to the
inner region of the halo (purple dash-dotted curve).

Finally, changing (increasing) the background expansion rate relative to $\Lambda $CDM again leads to less clustering of particles, and thus lower
density profiles in the halos (pink dash-dot-dot-dot curves).

It is worth noting that although the velocity-dependent force, fifth force,
and varying mass all contribute to suppressing the internal density of the
halo, the varying mass contributes most, while the fifth force contributes
least. This differs from the finding of \cite{bprs2010}, that the
velocity-dependence force is the determining factor, possibly due to
differences in the models and their treatments. The fact that the
velocity-dependent force dominates over the fifth force (which is different
from what we have seen for matter power spectrum and mass function) is not
surprising, for, as mentioned above, the fifth force is set to increase both
the kinetic and potential energies, two effects somehow cancelling each
other out.

Fig.~\ref{fig:pf_s2} displays the same results for the models S2 and S2a-d,
and we can see qualitatively similar but stronger trends. The notable thing
is that in this case suppressing the fifth force further lowers the inner
density of the halo, an indication that here the deepening of potential
dominates over the increase in kinetic energy (both are due to the fifth
force).

Above we have just discussed the results for one specific halo (the most
massive one), while what we are more interested in is the general behaviour.
For this we have selected 10 out of the most massive halos and computed
their average density profile. The results for simulations S1 and S1a-d are
shown in Fig.~\ref{fig:pfr_s1}, in which we have plotted the fractional change
of the averaged halo internal density with respect to the $\Lambda $CDM
result. From this figure we see again that overall the internal density in
the inner region of halos is lower in (full) coupled scalar field models
than in $\Lambda $CDM. Both the velocity-dependent force and varying mass
tend to decrease the inner density, and both the fifth force and the
modified (slower) background expansion help increase it. The effects from
the varying mass and modified background expansion are dominant while the
other two effects are minor, confirming our observed pattern from a single
halo (Fig.~\ref{fig:pf_s1}).

We see a similar result for the models S2 and S2a-d, as shown in Fig.~\ref{fig:pfr_s2}. A notable difference here, however, is that the
velocity-dependent force becomes more important than in S1. Indeed, it is as
dominant as the varying mass effect in the inner region of the halos, though
it is still subdominant in the outer region; this at least shows some
agreement with \cite{bprs2010}. Note again that the fifth force is the
least important amongst all the four effects, which means that using a
simulation with $\Lambda$CDM plus a Yukawa-type fifth force we would be
unable to catch the (probably) most significant effects from a coupled
scalar field.

\section{Summary and Conclusion}

\label{sect:con}

To summarise: in this paper, with the aid of $N$-body simulations, we have
investigated the different impacts of a coupled scalar field on cosmic
structure formation, and assessed their qualitative effects and quantitative
importance. The scalar field coupling influences the structure formation
mainly through a velocity-dependent force, a fifth force, a modification of
the particle mass (or the source of the Poisson equation) and a modified
background expansion rate. We have investigated how dropping each one of these
factors leaves imprints on the key structure formation observables, like
matter power spectrum, mass function and the internal density profile of
dark matter halos.

For the matter power spectrum and mass function, we find that the
modified background expansion rate is by far the most important effect that the
scalar-coupling can have, followed in turn by  the variation of
particle mass, fifth force and the velocity-dependent force. The cosmic
expansion becomes slower than that in $\Lambda $CDM due to the scalar field,
which means that structure has more time to form. The fifth force increases
the mutual attractive force between particles, strengthening the collapse of
overdense regions; both the fifth force and the velocity-dependent force
could speed up particles, making the collapse faster. As a result, all these
three effects help boost the growth of structure. On the other hand, the
source of the Poisson equation is decreased due to the coupling function $C(\varphi )$, resulting in weaker gravity and weakened structure formation.

The internal density profiles for dark matter halos are more interesting,
and as we have seen the combined effect of a scalar-coupling can be to
suppress the density of the inner regions of the halos (or to
distribute particles more towards outer regions). As one might expect, the
modified cosmic expansion rate is again the most important scalar-coupling
effect here, and with it dropped there will be less structure formation and lower
halo density profiles. The variation of particle masses, or the modification
to the source of the Poisson equation, is the second largest single effect,
followed by the velocity-dependent force and fifth force. Roughly speaking,
the particles tend to move towards (by a process of relaxation and
virialisation) the inner regions of halos if their kinetic energy is reduced
and/or the central potential gets deeper, and vice versa \citep{bprs2010}.
In this regard, the velocity-dependent force speeds up particles and the
variation of particle mass weakens the central potential, both in favour of
lower central densities in halos (cf.~Figs.~\ref{fig:pf_s1} - \ref{fig:pfr_s2}). The fifth force has two opposite effects -- to speed up
particles (and so increase the kinetic energy) and to deepen the total
potential. They cancel each other out to a certain extent, so making the
fifth force less influential in determining the density profiles (than the
velocity-dependent force)

Our result is marginally consistent with previous analyses (\emph{e.g.}, 
\cite{bprs2010}) of the halo density profiles, but shows discrepancy about
whether the velocity-dependent force is more important than the varying
particle mass or not. The difference might be due to different model
specifications and treatments.

One of our most important results is that in many cases the fifth force, which is the most
well-known consequence of a coupling between matter and a scalar field, is not
the most important in affecting the structure formation. The key point is
that, when we introduce such a coupling, other new effects are also brought
in, and these can often be much more influential. It is in this regard that
the advantages of full $N$-body simulations \cite{mqmab2004, bprs2010,
lz2009, lz2010, lb2010}, which take full account of all associated
effects, are increasingly significant.

\section*{Acknowledgments}

The work described in this paper has been performed on \texttt{COSMOS}, the
UK National Cosmology Supercomputer. The matter power spectrum and halo properties
are computed using \texttt{POWMES} \citep{cjnp2009} and \texttt{MHF} \cite{gkg2004}
respectively. We thank Marco Baldi for comments on the draft. B.~Li is supported by a Junior Research Fellowship at Queens'
College, Cambridge, and the Science and Technology Facility Council of the
United Kingdom.

\label{lastpage}

\begin{thebibliography}{}
\bibitem[\protect\citeauthoryear{Amendola}{2000}]{amendola2000} Amendola L., 2000, PRD62, 043511
\bibitem[\protect\citeauthoryear{Amendola}{2004}]{amendola2004} Amendola L., 2004, PRD69, 103524
\bibitem[\protect\citeauthoryear{Baldi}{2010}]{baldi2010} Baldi M., 2010, arXiv:1005.2188 [astro-ph.CO]
\bibitem[\protect\citeauthoryear{Baldi \& Pettorino}{2010}]{bp2010} Baldi M., Pettorino V., 2010, arXiv: 1006.3761 [astro-ph.CO]
\bibitem[\protect\citeauthoryear{Baldi \& Viel}{2010}]{bv2010} Baldi M., Viel M., 2010, arXiv: 1007.3736 [astro-ph.CO]
\bibitem[\protect\citeauthoryear{Baldi {\it et~al.}}{2010}]{bprs2010} Baldi M., Pettorino V., Robbers G., Springel V., 2010, MNRAS, 403, 1684
\bibitem[\protect\citeauthoryear{Bean \& Magueijo}{2001}]{bm2001} Bean R., Magueijo J., 2001, Phys.~Lett.~B17, 177
\bibitem[\protect\citeauthoryear{Bean {\it et~al.}}{2008}]{bflt2008} Bean R., Flanagan E.~E., Laszlo I., Trodden M., 2008, PRD78, 123514
\bibitem[\protect\citeauthoryear{Bean, Flanagan \& Trodden}{2008}]{bft2008} Bean R., Flanagan E.~E., Trodden M., 2008, PRD78, 023009
\bibitem[\protect\citeauthoryear{Boehmer {\it et~al.}}{2008}]{bclm2008} Boehmer C.~G., Caldera-Cabral G., Lazkoz R., Maartens R., 2008, PRD78, 023505
\bibitem[\protect\citeauthoryear{Boehmer {\it et~al.}}{2009}]{bcclm2009} Boehmer C.~G., Caldera-Cabral G., Chan N., Lazkoz R., Maartens R., 2008, PRD81, 083003
\bibitem[\protect\citeauthoryear{Caldera-Cabral, Maartens \& Schaefer}{2009}]{cms2009} Caldera-Cabral G., Maartens R., Schaefer B.~M., 2009, JCAP, 0907, 027
\bibitem[\protect\citeauthoryear{Colombi {\it et al}}{2009}]{cjnp2009} Colombi S., Jaffe A., Novikov D., Pichon C., 2009, MNRAS, 393, 511
\bibitem[\protect\citeauthoryear{Copland {\it et al}}{2006}]{cst2006} Copeland E., Sami M., Tsujikawa S., 2006, Int.~J.~Mod.~Phys.~D, 15, 1753
\bibitem[\protect\citeauthoryear{De Boni {\it et al}}{2010}]{ddempb2010} De Boni C., Dolag K., Ettori S., Moscardini L., Pettpromp V., Baccigalupi C., 2010, arXiv: 1008.5376 [astro-ph.CO]
\bibitem[\protect\citeauthoryear{Farrar \& Rosen}{2007}]{fr2007} Farrar G.~R., Rosen R.~A., 2007, PRL, 98, 171302
\bibitem[\protect\citeauthoryear{Gill, Knebe \& Gibson}{2004}]{gkg2004} Gill S.~P.~D., Knebe A., Gibson B.~K., 2004, MNRAS, 351, 399
\bibitem[\protect\citeauthoryear{Hellwing \& Juszkiewicz}{2009}]{hj2009} Hellwing W.~A., Juszkiewicz R., 2009, PRD80, 083522
\bibitem[\protect\citeauthoryear{Hellwing, Juszkiewicz \& van de Weygaert}{2010}]{hjv2010} Hellwing W.~A., Juszkiewicz R., van de Weygaet R., 2010, arXiv: 1008.3930 [astro-ph]
\bibitem[\protect\citeauthoryear{Hellwing, Knollmann \& Knebe}{2010}]{hkk2010} Hellwing W.~A., Knollmann S.~R., Knebe A., 2010, MNRAS, 408, L104
\bibitem[\protect\citeauthoryear{Kesden \& Kamionkowski}{2006}]{kk2006} Kesden M., Kamionkowski M., 2006, PRL, 97, 131303; PRD74, 083007
\bibitem[\protect\citeauthoryear{Keselman, Nusser \& Peebles}{2009}]{knp2009} Keselman J.~A., Nusser A., Peebles P.~J.~E., 2009, PRD80, 063517
\bibitem[\protect\citeauthoryear{Keselman, Nusser \& Peebles}{2010}]{knp2010} Keselman J.~A., Nusser A., Peebles P.~J.~E., 2009, PRD81, 063521
\bibitem[\protect\citeauthoryear{Khoury \& Weltman}{2004}]{kw2004} Khoury J., Weltman A., 2004, PRD69, 044026
\bibitem[\protect\citeauthoryear{Knebe, Green \& Binney}{2001}]{kgb2001} Knebe A., Green A., Binney J., 2001, MNRAS, 325, 845
\bibitem[\protect\citeauthoryear{Koivisto}{2005}]{koivisto2005} Koivisto T, 2005, PRD72, 043516
\bibitem[\protect\citeauthoryear{Lee, Liu \& Ng}{2006}]{lln2006} Lee S., Liu G.~-C., Ng K.~-W., 2006, PRD73, 083516
\bibitem[\protect\citeauthoryear{Li}{2010}]{li2010} Li B., 2010, arXiv:1009.1406 [astro-ph.CO]
\bibitem[\protect\citeauthoryear{Li \& Barrow}{2010}]{lb2010} Li B., Barrow J.~D., 2010, arXiv:1005.4231 [astro-ph.CO]
\bibitem[\protect\citeauthoryear{Li \& Zhao}{2009}]{lz2009} Li B., Zhao H., 2009, PRD80, 044027
\bibitem[\protect\citeauthoryear{Li, Mota \& Barrow}{2010a}]{lmb2010a} Li B., Mota D.~F., Barrow J.~D., 2010, arXiv:1009.1396 [astro-ph.CO]
\bibitem[\protect\citeauthoryear{Li, Mota \& Barrow}{2010b}]{lmb2010b} Li B., Mota D.~F., Barrow J.~D., 2010, arXiv:1009.1400 [astro-ph.CO]
\bibitem[\protect\citeauthoryear{Li \& Zhao}{2010}]{lz2010} Li B., Zhao H., 2010, PRD81, 104047
\bibitem[\protect\citeauthoryear{Linder \& Jenkins}{2003}]{lj2003} Linder E.~V., Jenkins A., 2003, MNRAS, 346, 573
\bibitem[\protect\citeauthoryear{Mainini {\it et~al.}}{2003}]{mmbk2003} Mainini R., Maccio A.~V., Bonometto S.~A., Klypin A., 2003, ApJ, 599, 24
\bibitem[\protect\citeauthoryear{Maccio {\it et~al.}}{2004}]{mqmab2004} Maccio A.~V., Quercellini C., Mainini R., Amendola L., Bonometto S.~A., 2004, PRD69, 123516
\bibitem[\protect\citeauthoryear{Mota \& Shaw}{2007}]{ms2007} Mota D.~F., Shaw D.~J., 2007, PRD75, 063501
\bibitem[\protect\citeauthoryear{Nusser, Gubser \& Peebles}{2005}]{ngp2005} Nusser A., Gubser S.~S., Peebles P.~J.~E., 2005, PRD71, 083505
\bibitem[\protect\citeauthoryear{Sandvik, Barrow \& Magueijo}{2002}]{sbm2002} Sandvik H., Barrow J.~D., Magueijo J., 2002, PRL88, 031302
\bibitem[\protect\citeauthoryear{Simpson, Jackson \& Peacock}{2010}]{sjp2010} Simpson, F., Jackson B.~M., Peacock J.~A., 2010, arXiv:1004.1920 [astro-ph.CO]
\bibitem[\protect\citeauthoryear{Springel \& Farrar}{2007}]{sf2007} Springel V., Farrar G.~R., 2007, MNRAS, 380, 911
\bibitem[\protect\citeauthoryear{Valiviita, Maartens \& Majerotto}{2010}]{vmm2010} Valiviita J., Maartens R., Majerotto E., 2010, MNRAS, 402, 2355
\bibitem[\protect\citeauthoryear{Zhao {\it et~al.}}{2010}]{zmlhf2010} Zhao H., Maccio A.~V., Li B., Hoekstra H., Feix M., 2010, ApJ, L712, 179
\end{thebibliography}
\end{document}